\documentclass[a4paper, 12pt]{article}
\usepackage[right=2.5cm, twoside, left=2.5cm, bottom=2.8cm]{geometry}
\usepackage{setspace}
\usepackage{graphicx}
\usepackage{amsmath}
\usepackage{amssymb}
\usepackage[english]{babel}
\usepackage{xcolor}
\usepackage[separate-uncertainty = true,multi-part-units=single]{siunitx}
\usepackage{nicefrac}
\usepackage[colorlinks=true, pdfstartview=FitV, linkcolor=blue,
citecolor=blue, urlcolor=blue]{hyperref}
\usepackage[capitalise]{cleveref}
\usepackage{authblk}
\begin{document}
\title{\textbf{Ultrafast Hole Spin Qubit with Gate-Tunable Spin-Orbit Switch Functionality}}
\author[1]{Florian N. M. Froning\thanks{These authors contributed equally.}}
\author[1]{Leon C. Camenzind$^*$}
\author[1,2]{Orson A. H. van der Molen}
\author[2]{Ang Li}
\author[2]{Erik P. A. M. Bakkers}
\author[1]{Dominik M. Zumb\" uhl\thanks{Author to whom correspondence should be addressed. Electronic mail: \texttt{floris.braakman@unibas.ch}.}}
\author[1]{Floris R. Braakman\thanks{Author to whom correspondence should be addressed. Electronic mail: \texttt{dominik.zumbuhl@unibas.ch}.}}

\affil[1]{University of Basel, Klingelbergstrasse 82, 4056 Basel, Switzerland}
\affil[2]{Department of Applied Physics, Eindhoven University of Technology, P.O. Box 513, 5600 MB Eindhoven, The Netherlands}

\maketitle
\clearpage

\begin{abstract}
Quantum computers promise to execute complex tasks exponentially faster than any possible classical computer, spurring breakthroughs in quantum chemistry, material science, and machine learning. However, quantum computers require fast and selective control of large numbers of individual qubits while maintaining coherence. Qubits based on hole spins in one-dimensional germanium/silicon nanostructures are predicted to experience an exceptionally strong yet electrically tunable spin-orbit interaction, allowing to optimize qubit performance by switching between distinct modes of ultrafast manipulation, long coherence, and individual addressability. Here, we use millivolt gate voltage changes to tune the Rabi frequency of a hole spin qubit in a germanium/silicon nanowire from 31 to 219\,MHz, its driven coherence time between 7 and 59\,ns, and its Land\'e g-factor from 0.83 to 1.27. We thus demonstrate spin-orbit switch functionality, with on/off ratios of roughly 7 in this first experimental implementation, which could be further increased through improved gate design. Finally, we use this control to optimize our qubit further and approach the strong driving regime, with spin-flipping times as short as $\boldsymbol{\sim}$1 ns.
\end{abstract}
\clearpage

Spin qubits defined in Si and Ge quantum dots are of particular interest for scaling up quantum circuits due to their small size, speed of operation, and compatibility with semiconductor industry~\cite{Zwanenburg2013,Kloeffel2013,Vandersypen2017,Scappucci2020}. Both materials feature a low natural abundance of non-zero nuclear spins, which has led to the demonstration of long qubit coherence times~\cite{Zwanenburg2013,Tyryshkin2012,Yoneda2018}, as well as single-~\cite{Kawakami2014,Veldhorst2014,Yoneda2018} and two-qubit~\cite{Veldhorst2015,Zajac2018,Watson2018,Hendrickx2020} operations with high fidelity. Most of this research has been performed using electron spin states defining the qubit~\cite{Loss1998}. Hole spin qubits~\cite{Bulaev2005,Maurand2016,Crippa2019,Scappucci2020} have recently gained attention since they potentially enable faster quantum operations and a higher level of control over qubit parameters~\cite{Kloeffel2011,Maier2013,Kloeffel2013a,Kloeffel2018}. In addition, hole spins in Ge and Si may have improved relaxation and decoherence times, since they do not exhibit a valley degeneracy and their wave function has reduced overlap with nuclear spins~\cite{Yang2013,Prechtel2016}. Importantly, spin-orbit interaction (SOI) can be exceptionally strong for hole spins in low-dimensional nanostructures~\cite{Durnev2014,Marcellina2017}, particularly in Ge- or Si-based nanowires~\cite{Kloeffel2011,Kloeffel2018}. 
This enables very fast spin control through electric-dipole spin resonance (EDSR)~\cite{Golovach2006,Nowack2007,Bulaev2007,vandenBerg2013}, where a time-varying electric field periodically displaces the hole wave function, thus creating an effective periodic magnetic field through the SOI. In this way, EDSR can be used for all-electrical spin manipulation without requiring micromagnets~\cite{Pioro-Ladriere2008} or co-planar striplines~\cite{Koppens2006}, which add to device complexity.

Rabi frequencies of around 100\,MHz have been measured for hole spins~\cite{Watzinger2018,Hendrickx2020}, but predictions for one-dimensional systems range even up to 5\,GHz, made possible by the particularly strong direct Rashba spin-orbit interaction~\cite{Kloeffel2011,Kloeffel2013a}. Conversely, this strong SOI may lead to an undesired enhancement of qubit relaxation and dephasing rates, via coupling to phonons or charge noise. However, the direct Rashba SOI is also predicted to be tunable to a large extent through local electric fields~\cite{Kloeffel2011,Maier2013,Kloeffel2018}, enabling electrical control over the SOI strength and Land\'e $g$-factor. Such electrical tunability provides a path towards a spin qubit with switchable interaction strength, using what we term a \emph{spin-orbit switch}. The spin-orbit switch can be used to selectively idle a qubit in an isolated configuration of weak SOI and low decoherence (\textit{Idle}-state), while for fast manipulation it is tuned to a regime of strong SOI (\textit{Control}-state) and is selectively coupled to an EDSR driving field or microwave resonator by controlling the qubit Zeeman energy~\cite{Kloeffel2013a,Burkard2020}.
\noindent Here, we experimentally realize the key components of this approach, through the demonstration of an ultrafast and electrically tunable hole spin qubit in a Ge/Si core/shell nanowire. We use SOI-mediated EDSR to perform fast two-axis qubit control and implement Ramsey and Hahn echo pulsing techniques to compare the qubit's coherence times. We then demonstrate a high degree of electrical control over the Rabi frequency, $g$-factor, and driven qubit decay time by tuning the voltage on one of the dot-defining gates, illustrating the basic ingredients of a spin-orbit switch. The spin-orbit switch functionality that we demonstrate here shows moderate on/off ratios of about 7 for both Rabi frequency and coherence times, which could in future devices likely be increased through improved gate design. We extract a spin-orbit length that is extraordinarily short and electrically tunable over a large range down to \,3\,nm for holes of heavy-hole mass. This control allows us to optimize our qubit for speed of operation, resulting in Rabi frequencies as large as 435\,MHz.

\section{Setup and Measurement Techniques}

\begin{figure}
	\centering
	\includegraphics{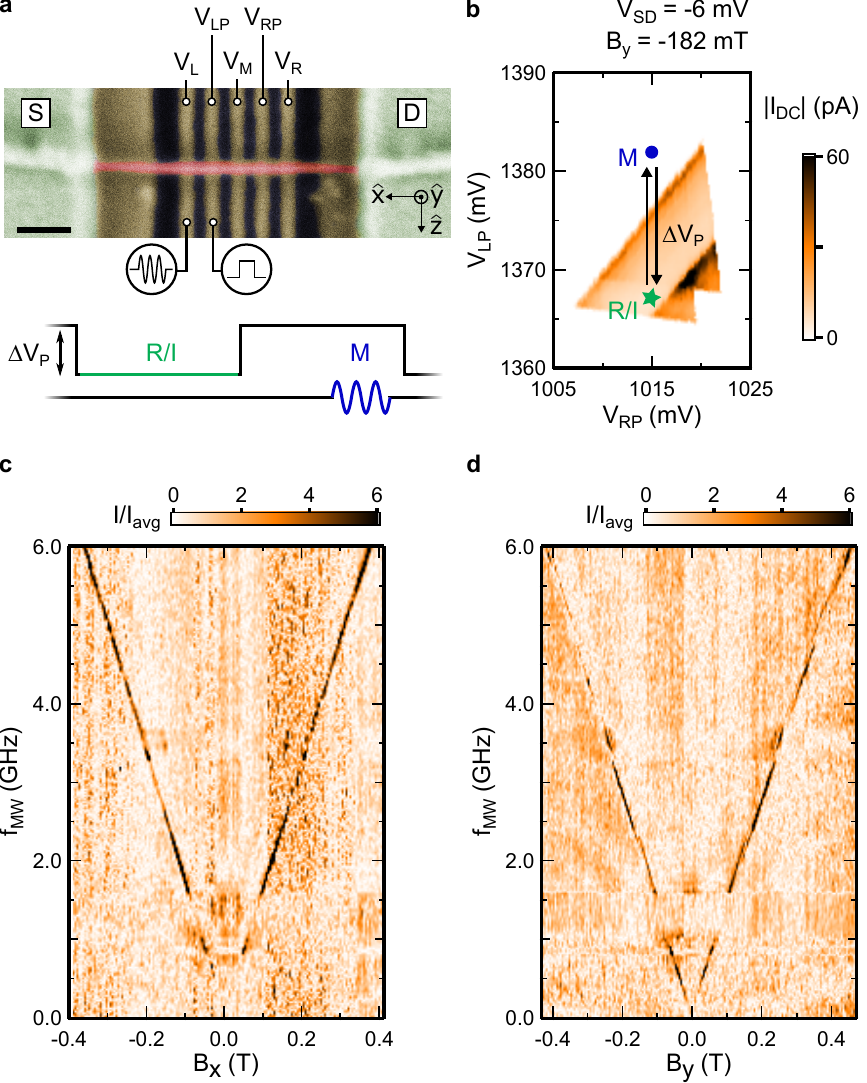}
	\caption[Setup and PSB and V]{\textbf{Experimental setup and electric dipole spin resonance.} \textbf{a} Scanning electron micrograph of a cofabricated device, showing source (S) and drain (D) contacts and gates, as labeled. The scalebar corresponds to 100\,nm. The inset at the bottom illustrates the pulse scheme. The points R, I and M indicate the locations of the readout, initialization, and manipulation stages, respectively, of the pulsing scheme (see \textbf{b}). The depth of the square pulse is~$\Delta V_\text{P}$. \textbf{b} Measurement of a set of bias triangles taken with a source-drain voltage $V_\text{SD}=\,-6\,$mV showing Pauli spin blockade which is partially lifted at a finite magnetic field~$B_\text{y}=\,-182\,$mT. \textbf{c, d} Spin blockade leakage current indicating electric dipole spin resonance as a function of microwave frequency and magnetic field magnitude in $\hat{x}$ (\textbf{c}) and $\hat{y}$ (\textbf{d}) direction. Horizontal bands of decreased intensity are due to microwave resonances in the high-frequency circuitry. For detailed measurement parameters and description of data analysis, see Methods.}
	\label{fig:figure1}
\end{figure}

Figure~1\,\textbf{a} shows a scanning electron micrograph of the device comprising five gates beneath a Ge/Si core/shell nanowire~\cite{Brauns2016b,Conesa-Boj2017,Froning2018}. A depletion-mode few-hole double quantum dot is formed inside the nanowire by positively biasing the five bottom gates. Throughout this work, we perform measurements of electronic transport through the double quantum dot, using the source (S) and drain (D) contacts indicated in Figure~1\,\textbf{a} (for more details about the device and measurement setup, see Methods). We operate the device at a transition exhibiting Pauli spin blockade~\cite{Ono2002}, which we use for spin readout in transport measurements. 

In our setup, gates L and LP are connected via bias-tees to high-frequency lines as indicated in Figure~1\,\textbf{a}, allowing us to apply square voltage pulses and microwave bursts to these gates. The measurements are performed with a two-stage pulse scheme (see inset Fig.~1\,\textbf{a}). First, the system is initialized at point I (see Fig.~1\,\textbf{b}) in a spin-blockaded triplet state. Then, with a square pulse of depth $\Delta V_\text{P}$, it is pulsed into Coulomb blockade to point M where a microwave burst of duration $t_\text{burst}$ is applied. Finally, back at the readout point R, a current signal is measured if the spins were in a singlet configuration after manipulation. 

Figures~1\,\textbf{c} and \textbf{d} show typical EDSR measurements, where the microwave frequency $f_\text{MW}$ is swept versus the applied magnetic field  $\vec{B}_\text{ext}$ along the $\hat{x}$- and $\hat{y}$-axis, respectively. On resonance, the spin is rotated, lifting spin blockade and leading to an increased current. From Figures~1\,\textbf{c} and \textbf{d}, we extract $g_\text{x}=1.06$ and $g_\text{y}=1.02$. With $\vec{B}_\text{ext}$ aligned along the $\hat{z}$ direction, no EDSR signal could be observed, as will be discussed later.

\section{Coherent Manipulation and Two-Axis Control}
\begin{figure}
	\centering
	\includegraphics{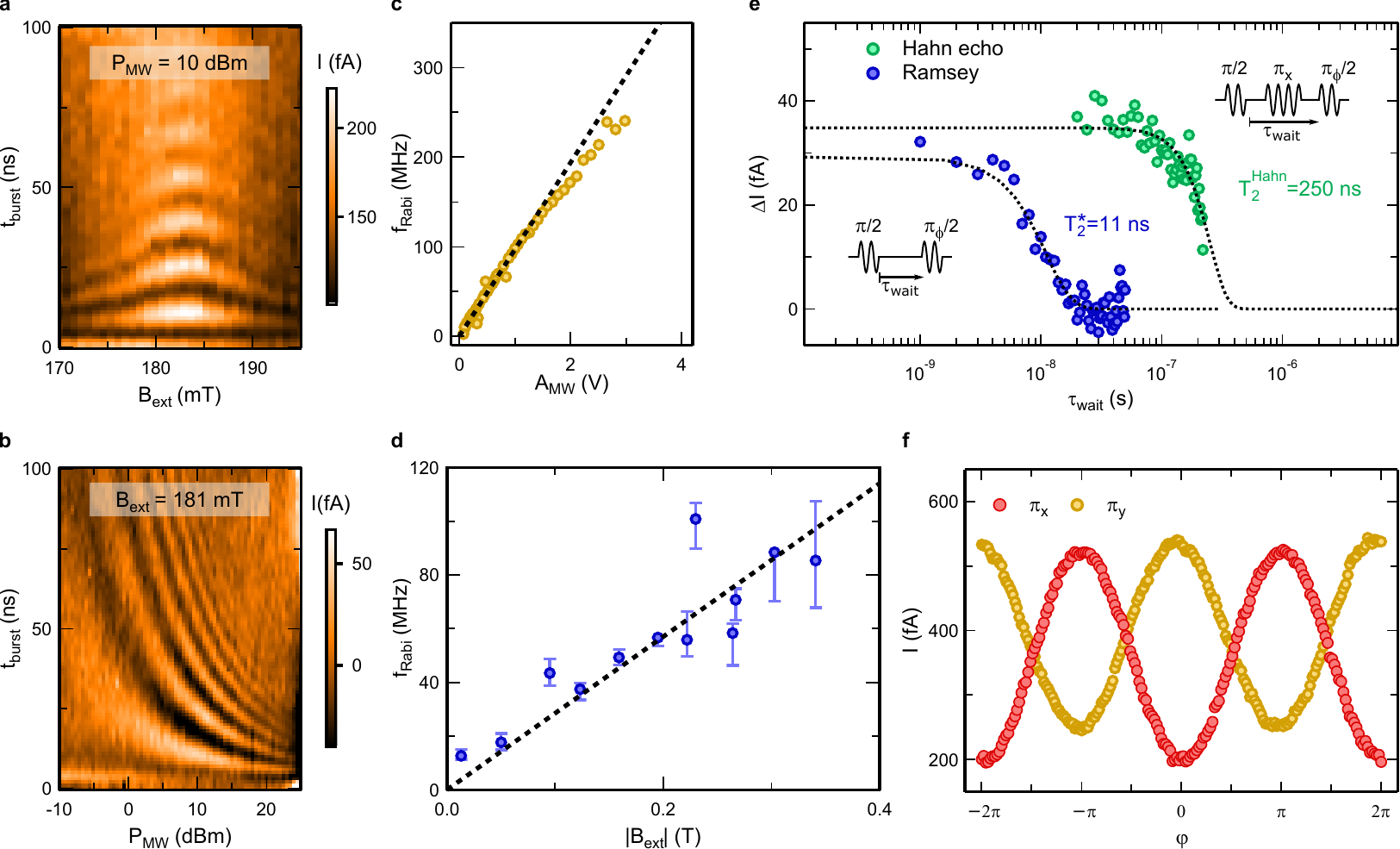}
	\caption[Coherent manipulation, SOI and power dependence]{\textbf{Coherent qubit control.} \textbf{a} Measurement of the current as a function of microwave burst duration and magnetic field. We observe a Rabi frequency of 72\,MHz. \textbf{b} Power dependence of Rabi oscillations in the same configuration as in~\textbf{a}. \textbf{c} Extracted Rabi frequency as a function of the microwave amplitude, from fits of the data in~\textbf{b} (see Methods for details). The black dashed line is a linear fit to the extracted Rabi frequencies. \textbf{d} Rabi frequency as a function of the magnitude of the external magnetic field. The black dashed line is a linear fit to the data over the whole range with zero offset. The error bars correspond to the inaccuracy of the frequency-dependent power calibration (see Section~1 of the Supplementary Information). \textbf{e} Decay of Ramsey fringes (blue points) and Hahn echo (green points) as a function of the waiting time $\tau_\text{wait}$ between the two $\pi/2$-pulses. Insets show pulse sequences used for Ramsey (bottom left) and Hahn echo (top right). Black dotted lines are fits of the data to exponential decay. \textbf{f} Demonstration of two-axis qubit control by applying a Hahn echo sequence with two orthogonal $\pi$-pulses. The amplitudes of the fringes of the two datasets differ due to an offset in the calibration of the $\pi/2$-pulse duration between the two measurements.
	}
	\label{fig:figure2}
\end{figure}

To demonstrate coherent control, we now vary the pulse duration $t_\text{burst}$ and observe Rabi oscillations, in the form of the typical chevron pattern shown in Figure~2\,\textbf{a}. Figure~2\,\textbf{b} shows the dependence on the microwave power $P_\text{MW}$. From line cuts, we extract the Rabi frequency $f_\text{Rabi}$ (see Methods), which is shown in Figure~2\,\textbf{c} as a function of the microwave amplitude. The data at low amplitudes is in good agreement with a linear fit (black dashed line in Fig.~2\textbf{c}), as expected theoretically. The saturation behaviour at higher amplitudes likely originates from smaller effective displacement due to anharmonicity~\cite{Yoneda14,Takeda16} in dot confinement for the particular gate voltage configuration used here, leading to sub-linear dependence on the amplitude $A_\text{MW}$ of the microwave driving field. Additionally, the strong and tunable SOI in our system could potentially lead to such non-linearities.\\

In the presence of SOI, the oscillating electric field on gate $V_\text{LP}$ due to the microwaves gives rise to an oscillating effective magnetic field $\vec{B}_\text{eff}(t)$, with magnitude~\cite{Golovach2006}:
\begin{equation}
|\vec{B}_\text{eff}(t)| = 2 |\vec{B}_\text{ext}|\cdot \frac{l_\text{dot}}{l_\text{so}}\cdot\frac{e |\vec{E}_\text{MW}(t)|l_\text{dot}}{\Delta_\text{orb}},
\label{eq:GolovachFormula}
\end{equation}
with $e$ the elementary charge, $\vec{E}_\text{MW}(t)$ the ac electric field in the dot generated by the microwaves, $l_\text{dot}$ the dot length, $\Delta_\text{orb}\propto l_\text{dot}^{-2}m_\text{eff}^{-1}$ the orbital level splitting, and $l_\text{so}$ the spin-orbit length, which we define here as setting the distance a hole has to travel along the nanowire to have its spin flipped due to SOI. 
This effective field $\vec{B}_\text{eff}$ drives the Rabi oscillations, with Rabi frequency $f_\text{Rabi}=g_\perp\mu_\text{B}|\vec{B}_\text{eff}(t)|/2h$, with $g_\perp$ the $g$-factor along the direction of $\vec{B}_\text{eff}$ and thus perpendicular to $\vec{B}_\text{ext}$.
From equation~\eqref{eq:GolovachFormula} we see that $|\vec{B}_\text{eff}|$ scales linearly with $|\vec{B}_\text{ext}|$. We measure the Rabi frequency for different $|\vec{B}_\text{ext}|$ and plot the result in Figure~2\,\textbf{d}.
Despite relatively large error bars at higher fields due to the inaccuracy of the frequency-dependent microwave power calibration (see Section~1 of the Supplementary Information), the measurement agrees well with a linear dependence of the Rabi frequency on $|\vec{B}_\text{ext}|$, as expected for SOI-mediated EDSR~\cite{Golovach2006,Nowack2007}.

Next, in order to characterize the free induction decay, we apply a Ramsey pulse sequence, as depicted in Figure~2\,\textbf{e}. A fit to a Gaussian decay yields the dephasing time~$T_2^*=\,11\pm 1$\,ns. This value is one order of magnitude smaller than in comparable hole spin qubit systems~\cite{Higginbotham2014,Hendrickx2020,Watzinger2018}. This may be attributed to low-frequency noise, which could for instance be due to gate voltage fluctuations, frequency jitter of the microwave source, charge fluctuators, or residual nuclear spin noise. Nevertheless, we can mitigate this to a large extent using a Hahn echo sequence, prolonging coherence by a factor of $\sim$25, thus demonstrating efficient decoupling of the qubit from low-frequency noise. In our measurements, we find no clear indication of decay due to spin relaxation. Indeed, previous experimental~\cite{Hu2012} and theoretical~\cite{Maier2013} works have found spin relaxation times in Ge/Si nanowires to be in the millisecond to second regime, much longer than can be probed using our pulsing and read-out scheme.

Finally, we use a modified Hahn echo pulse sequence to demonstrate two-axis control. We employ either a $\pi_\text{x}$- or a $\pi_\text{y}$-pulse and vary the phase of the second $\pi_\phi/2$-pulse (see schematics in Fig.~2\,\textbf{e}). This results in two sets of Ramsey fringes as shown in Figure~2\,\textbf{f}, which are phase-shifted by $\pi$. These measurements demonstrate universal, two-axis control of the hole spin qubit.

\section{Spin-Orbit Switch Functionality}
\begin{figure}
	\centering
	\includegraphics[width=\textwidth]{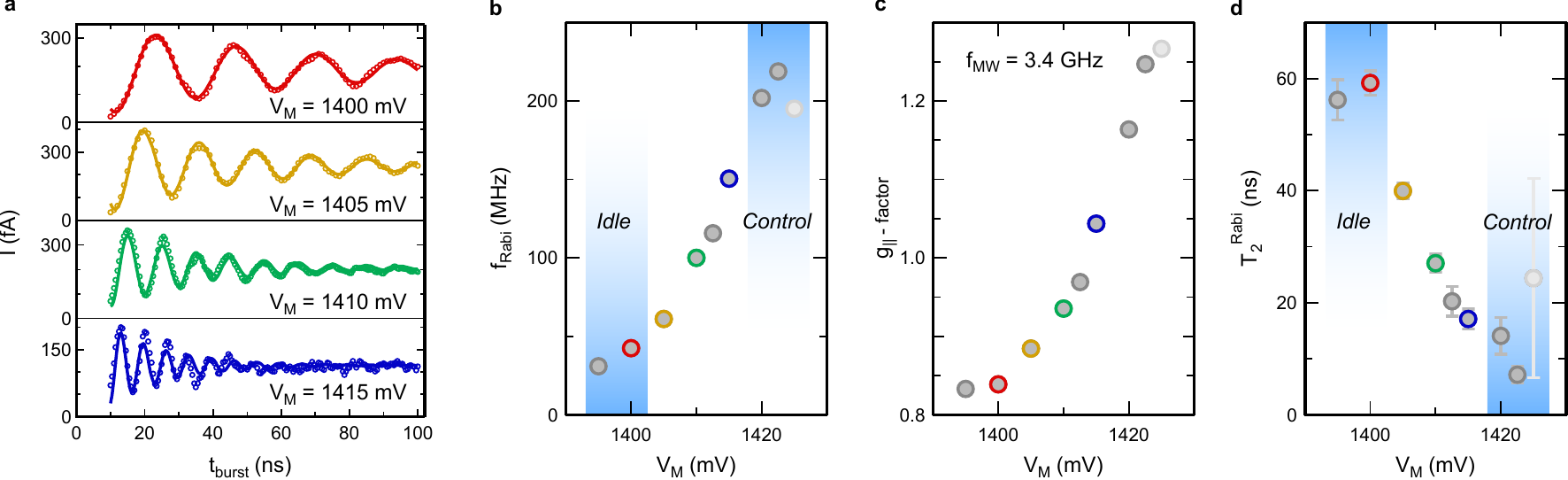}
	\caption[Tunability and B rotation]{\textbf{Electrical tunability of qubit parameters.} \textbf{a} Rabi oscillations for four different gate voltage values $V_\text{M}$. Here, all parameters were kept constant, except when indicated in the plots.
		\textbf{b, c, d} Rabi frequency, $g$-factor, and $T_2^\text{Rabi}$ as a function of the gate voltage $V_\text{M}$, as extracted from fits to line cuts such as shown in \textbf{a}. Error bars correspond to the standard deviations resulting from the fitting. Insets in \textbf{b} and \textbf{d} indicate possible idling and control points. These points define a spin-orbit switch with on/off ratio of 7.1 and 8.3 for the $f_\text{Rabi}$ and $T_2^\text{Rabi}$, respectively. Note that the measurement corresponding to the point at $V_\text{M}=\,1425$\,mV (light grey) suffers from a low signal-to-noise ratio, due to reduced interdot tunnel coupling, resulting in a large uncertainty in the analysis of the data at this point.
	}
	\label{fig:figure3}
\end{figure}
The measurements of Figure~2 establish Ge/Si nanowires as a platform for hole spin qubits. The particular direct Rashba SOI~\cite{Kloeffel2011,Kloeffel2018} provides a unique way to electrically control the qubit via the SOI strength and qubit Zeeman energy~\cite{Maier2013,Kloeffel2013a}. This tunability can be exploited for optimizing qubit relaxation and dephasing times, as well as selective coupling of the qubit to EDSR drive fields or microwave resonators~\cite{Kloeffel2013a,Trif2008,Nigg2017}. Here, we demonstrate this distinct gate-tunability of hole spin qubits in Ge/Si core/shell nanowires, where we investigate electrical control over the $g$-factor, Rabi frequency, and coherence time.

The gate voltages not only provide the electrostatic confinement but also constitute a static electric field on the order of tens of~V/$\mathrm{\mu}$m inside the quantum dots, which has a significant effect on the strength of SOI~\cite{Kloeffel2018,Kloeffel2011}. Figure~3\,\textbf{a} shows example Rabi oscillations for four different gate voltages $V_\text{M}$. Here, $f_\text{MW}$ and $P_\text{MW}$ are kept fixed, while $|\vec{B}_\text{ext}|$ is adjusted to compensate for changes in the $g_\parallel$-factor along $B_\text{ext}$, keeping the qubit on resonance with the microwave drive. As shown in Figure~3\,\textbf{b}, we find that the Rabi frequency depends strongly on $V_\text{M}$, with a gate voltage change of 30\,mV resulting in a 7-fold increase of the Rabi frequency. 

For SOI-mediated spin rotations~\cite{Golovach2006}, the Rabi frequency is proportional to the effective magnetic field given by Eq.~\eqref{eq:GolovachFormula} and the $g$-factor $g_\perp$. Therefore, the Rabi frequency depends on the spin-orbit length $l_\text{so}$, the ac electric field $|\vec{E}_\text{MW}(t)|$ created through the periodic gate voltage modulation, $\vec{B}_\text{ext}$, the quantum dot confinement $\Delta_\text{orb}$, and the $g$-factor. Despite an observed change of $g_\parallel$ with $V_\text{M}$~\cite{Dmytruk2018,Maier2013} by a factor of~1.5 (see Fig.~3\,\textbf{c}), the effect on the Rabi frequency is small: at constant Zeeman energy, we can write $f_\text{Rabi} \propto f_\text{MW} \cdot g_\perp/g_\parallel $. Hence, if the $g$-factor anisotropy $g_\perp / g_\parallel$ is only weakly affected by gate voltages, as observed here (see Supplementary Information Figs.~S3\,\textbf{d, e}), then the Rabi frequency change is correspondingly small.

We have carefully analyzed each of the contributions to the change of the Rabi frequency (see Section~2.2 of the Supplementary Information). In particular, we find that the orbital level splitting $\Delta_\text{orb}$ shows only a weak dependence on gate voltage $V_\text{M}$ and that the electric field amplitude $E_\text{MW}$ stays roughly constant. These effects are not sufficient to explain the large change in $f_\text{Rabi}$ and we therefore find that the large change must mostly be attributed to a gate-tunability of the spin-orbit length $l_\text{so}$. Using equation~\eqref{eq:GolovachFormula}, we extract upper bounds of $l_\text{so}$ (see Supplementary Information Section~2.2.1). We find remarkably short values of $l_\text{so}$ that are tuned from 28\,nm down to 3\,nm. Here we assume a heavy-hole effective mass, as suggested by independent transport measurements at high magnetic field~\cite{Froning2020}. Such a strong SOI was predicted for the direct Rashba SOI~\cite{Kloeffel2011,Kloeffel2018}. This range of $l_\text{so}$ overlaps with values found in antilocalization~\cite{Higginbotham2014a} and spin blockade experiments~\cite{Froning2020}. Finally, while the direct Rashba SOI term is predicted to be very strong in this system, additional weaker SOI terms may also be present, but cannot be distinguished here.

Besides the Rabi frequency, also the coherence is strongly affected by $V_\text{M}$, as shown in Figure~3\,\textbf{a}. We plot the characteristic driven decay time $T^\text{Rabi}_2$ in Figure~3\,\textbf{d}, finding that it scales roughly inversely with $f_\text{Rabi}$ and $g_\parallel$: a short decay time coincides with a high Rabi frequency, and vice versa. Together with the tunability of the Rabi frequency, this control over the qubit coherence time allows us to define (see insets Figs.~3\,\textbf{b} and \textbf{d}) a fast qubit manipulation point (\textit{Control}) and a qubit idling point featuring significantly improved coherence (\textit{Idle}). This demonstrates the functionality of a spin-orbit switch, though here with modest on/off ratios for the switching of $f_\text{Rabi}$ and $T_2^\text{Rabi}$ between the \textit{Control} and \textit{Idle} points.

Moreover, the variation of $g_\parallel$ in Figure~3\,\textbf{c} effectively adds a third mode of operation to the spin-orbit switch, where individual qubits can be selectively tuned, for instance in and out of resonance with a microwave cavity, enabling a switch for qubit-resonator coupling~\cite{Trif2008,Nigg2017}. Finally, we find that the pulse depth $\Delta V_\text{P}$ can also be used to tune of $f_\text{Rabi}$ and $g_\parallel$ (see Section~2.1 of the Supplementary Information), indicating that dynamically pulsing these quantities is feasible.

\section{Ultrafast Rabi Oscillations}
\begin{figure}
	\centering
	\includegraphics{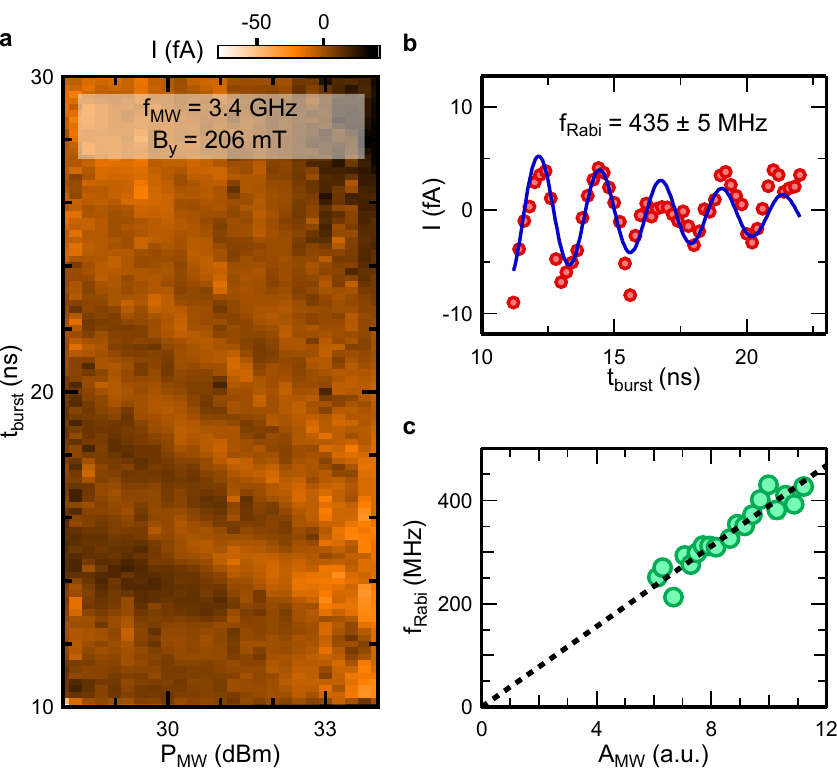}
	\caption[Fast Rabi]{\textbf{Ultrafast coherent control.} \textbf{a} Power-dependence of ultrafast Rabi oscillations. \textbf{b} Line cut of data shown in \textbf{a} at a microwave power of 34\,dBm. The data is fitted (blue solid curve) as in Fig.~2, after subtraction of a linear background. \textbf{c} Rabi frequency as a function of microwave amplitude, extracted from fits to line cuts in \textbf{a}. Dashed black line is a linear fit to the data.
	}
	\label{fig:figure4}
\end{figure}

In a next step, we now use the electrical tunability to optimize the gate voltages for a high Rabi frequency and furthermore increase the applied microwave power. In Figure~4, we show a measurement of ultrafast Rabi oscillations, with the maximum Rabi frequency reaching a value of $\sim$\,435\,MHz (see Fig.~4\,\textbf{b}), allowing for spin-flip times of the qubit as short as 1.15\,ns. As can be seen in Figure~4\,\textbf{c}, the Rabi frequency scales linearly with applied microwave amplitude in this regime of ultrafast qubit operation and shows no signs of saturation for the gate configuration used here, in contrast to Figure~2\,\textbf{b}. This indicates that even higher Rabi frequencies may be possible through the application of a higher microwave power. Note that pulse imperfections play a larger role for shorter pulse duration and higher amplitudes, which likely partially explains the decrease in $T_2^\text{Rabi}$ with increased microwave amplitude. 

\par Notably, the observed Rabi frequencies of over 400\,MHz are roughly~$1/8$ of the Larmor precession frequency of 3.4\,GHz. The system is thus approaching the strong driving regime where the rotating wave approximation is not applicable anymore, opening the possibility for ultrafast, non-sinusoidal spin-flipping~\cite{Kato2003,Laucht2016} that has not been realized before with conventional spin qubits. We note that in our experiment, effects of strong driving~\cite{Laucht2016} could contribute to the reduced visibility of Rabi oscillations at the high Rabi frequencies shown in the measurements of Figure~4.

\section{Conclusions}

We have demonstrated ultrafast two-axis control via EDSR of a hole spin qubit in a Ge/Si core/shell nanowire. Our measurements firmly demonstrate the feasibility of single-spin qubit operations on nanosecond timescales. Ideally, such fast operations would be combined with long qubit coherence times. 
We observe a relatively short inhomogeneous dephasing time, which is likely related to technical pulsing challenges at such short timescales. This may be resolved with improved instrument control. Also, we measure a much larger spin echo decay time, which indicates the presence of low-frequency noise affecting our qubit. Finally, the use of a charge sensor will allow to decouple the quantum dots from the neighboring Fermi reservoirs, likely leading to a significant further enhancement of the coherence time.

We have demonstrated a 7-fold increase of the Rabi frequency for a relatively small change in gate voltage. Similarly, we find that the driven decay time of our qubit can be tuned by the same gate voltage, demonstrating the working principle of a spin-orbit switch. Thus far, the spin-orbit switch is limited to moderate on/off ratios of $f_\text{Rabi}$ and $T^\text{Rabi}_2$. However, improved devices with gates designed for precise engineering of the electric field profile could in future experiments lead to a higher level of control over the SOI, resulting in higher on/off ratios as suggested by theoretical work~\cite{Kloeffel2013a}.
Our measurements indicate the presence of an exceptionally strong spin-orbit interaction in Ge/Si core/shell nanowires, in qualitative agreement with predictions of direct Rashba SOI~\cite{Kloeffel2011,Kloeffel2018}. A more quantitative comparison to theory, as well as improved gate switching, requires precise engineering of the electric field and single-hole dot occupation, both of which can be achieved through optimization of the gate design.

The high tunability of the qubit demonstrates the suitability of the platform for the implementation of a qubit with switchable interaction strengths. The effect of the gate voltages and the pulse depth on the qubit resonance frequency and the Rabi frequency have the potential to dynamically pulse the characteristic qubit parameters and interaction strengths from a qubit manipulation to an idling point. Furthermore, the spin-orbit switch could allow tuning to `sweet spots' of operation, where the SOI strength is to first order insensitive to charge noise, leading to enhancement of qubit coherence~\cite{Kloeffel2018}. Finally, the strong spin-orbit interaction holds potential for realizing fast entangling operations between distant spin qubits, mediated by a microwave resonator~\cite{Trif2008,Nigg2017,Kloeffel2013a,Borjans2019,Burkard2020}.

\section{Methods}
\subsection{Device Fabrication}
The device features a set of five gates with a width of \SI{20}{nm} and a pitch of \SI{50}{nm} defined by electron beam lithography on a $p^{++}$-doped Si chip covered with \SI{290}{nm} of thermal oxide. The gates are covered by a \SI{20}{nm} thick layer of Al$_\text{2}$O$_\text{3}$ grown by atomic layer deposition in order to electrically insulate them from the nanowire. A single Ge/Si core/shell nanowire with a core radius of about \SI{10}{nm} and a shell thickness of \SI{2.5}{nm}~\cite{Conesa-Boj2017} is placed deterministically across the set of gates using a micromanipulator. The nanowire is roughly aligned with the coordinate system in Figure~\ref{fig:figure1}\,\textbf{b} but the exact angle in the $\hat{x}\hat{z}$ plane is unknown. Finally, ohmic contacts are fabricated by electron beam lithography and metallized with Ti/Pd following a short dip in hydrofluic acid to remove the native oxide. The scanning electron micrograph shown in Fig.~\ref{fig:figure1}\,\textbf{a} is from a similarly fabricated device as used here.

\subsection{Experimental Setup}
The sample is wire-bonded to a printed-circuit board providing dc wiring and RF lines, coupled via bias tees. The circuit board is mounted in a Bluefors dilution refrigerator with a base temperature around $T_\text{base}=\SI{10}{mK}$, at which tempreature all measurements are taken. Each high-frequency line includes attenuators with combined values of $\sim$\SI{30}{dB}. A Basel Precision Instruments LNHR DAC is used to supply the dc voltages, and a Basel Precision Instruments LNHS I/V converter is used for readout of the qubit in transport. 

A Tektronix 7122C or AWG5208 arbitrary waveform generator is used to generate the square voltage pulses applied to gate $V_\text{LP}$. To drive the qubit, either an analog Keysight E8257D signal generator or a E8267D vector signal generator supplies the microwave tone. For measurements at high microwave power a RF-Lambda model RFQ132070 amplifier was used. Two different configurations of the setup are used for microwave burst generation. For the measurements in Figures~\ref{fig:figure1}\,\textbf{c-d}, \ref{fig:figure2}\,\textbf{d}, \ref{fig:figure3}, and \ref{fig:figure4}, the amplitude of the microwaves is modulated by means of an RF-switch (ZASWA-2-50DRA+ from MiniCircuits), triggered by the arbitrary waveform generator. The RF-switch has a minimum pulse width of \SI{10}{ns}. For the measurements in Figures~\ref{fig:figure1}\,\textbf{b}, \ref{fig:figure2}\,\textbf{a-c, e, f}, the microwave bursts are generated by IQ modulation of the vector signal generator's microwave tone. Here, the minimum pulse width is \SI{6}{ns}.
In either configuration, a lock-in amplifier is used to chop the bursted microwaves at a frequency of \SI{89.75}{\hertz} and the I/V converter output is demodulated at his frequency. This allows us to separate the current signal due to the applied microwaves from the background.

\subsection{Data Analysis}
Rabi frequencies are extracted from fits to $ I\left(t_\text{burst}\right)=I_0 + C\cdot \sin\left(2\pi f_\text{Rabi}t_\text{burst}+\phi\right)\cdot\exp\left(-t_\text{burst}/T_2^\text{Rabi}\right)$. Here, $I_0$ is an offset, $C$ the amplitude, $\phi$ a phase shift, and  $T_2^\text{Rabi}$ the characteristic decay time. Furthermore, we post-processed raw data sets in the following ways. The data in Fig.~\ref{fig:figure1}\,\textbf{c} (\ref{fig:figure1}\,\textbf{d}) was offset by \SI{10}{mT} (\SI{20}{mT}) to compensate for trapped magnetic flux. Furthermore, the average value has been subtracted from each column and row of the raw data. Then each row has been divided by the average row value. Similarly, for the plots of Fig.\ref{fig:figure2}\,\textbf{b} and \ref{fig:figure4}\,\textbf{a}, the average value has been subtracted from each column and row of the raw data. In Fig.~\ref{fig:figure4}\,\textbf{a}, data for microwave burst times below the minimum pulse width achievable by our electronics is omitted.

\subsection{Measurement Details}
In the following we list the relevant parameters that were used for the various measurements.
For the measurements of Figs.~\ref{fig:figure1}\,\textbf{c-d}, a fixed pulse amplitude~$\Delta V_\text{P} = \SI{0.55}{V}$ and a burst duration~$t_\text{burst} = \SI{15}{ns}$ was used. In Figs.~\ref{fig:figure2}\,\textbf{a-c}, $\vec{B}_\text{ext}$ was oriented along the $-\hat{y}$-axis. For Fig.~\ref{fig:figure2}\,\textbf{d}, $f_\text{MW}=\SI{3.4}{GHz}$ was used and $\vec{B}_\text{ext}$ was oriented in the $\hat{x}\hat{y}$-plane, making an angle of \SI{40}{\degree} with the $\hat{y}$-axis. 
In Fig.~\ref{fig:figure2}\,\textbf{e}, the duration of the $\pi$-pulse ~$t_\pi=\SI{13}{ns}$, $P_\text{MW}=\SI{3}{dBm}$, $f_\text{MW}=\SI{2.6}{GHz}$, and $|\vec{B}_\text{ext}|=\SI{181}{mT}$ along the $-\hat{x}$-axis. For Fig.~\ref{fig:figure2}\,\textbf{f}, we used $P_\text{MW}=\SI{14}{dBm}$, $f_\text{MW}=\SI{3.4}{GHz}$ and $|\vec{B}_\text{ext}|=\SI{292}{mT}$, along the same direction as used for Fig.~\ref{fig:figure2}\,\textbf{d}. Finally, for the measurements of Fig.~\ref{fig:figure3}, we used $P_\text{MW}=\SI{25}{dBm}$ and the orientation of $|\vec{B}_\text{ext}|$ was the same as in Fig.~\ref{fig:figure2}\,\textbf{d}. 

For completeness, we also mention the other gate voltages used for the measurements of Fig.~\ref{fig:figure3}: $V_\text{L} = \SI{3710}{mV}$ and $V_\text{R} = \SI{1495}{mV}$, $V_\text{LP}$ and $V_\text{RP}$ depend on $V_\text{M}$, but are similar to the values used for Fig.~\ref{fig:figure1}\,\textbf{b}.

\section*{Data availability}
The data supporting the plots of this paper are available at the Zenodo repository with the following DOI: 10.5281/zenodo.4290131

\section*{Acknowledgments}
We thank Stefano Bosco, Bence Het\'enyi, Christoph Kloeffel, Daniel Loss, Arne Laucht, and Alex Hamilton for useful discussions. Furthermore, we acknowledge Sascha Martin and Michael Steinacher for technical support. This work was partially supported by the Swiss Nanoscience Institute (SNI), the NCCR QSIT, the NCCR SPIN, the Georg H. Endress Foundation, Swiss NSF (grant nr. 179024), the EU H2020 European Microkelvin Platform EMP (grant nr. 824109), and FET TOPSQUAD (grant nr. 862046).

\section*{Author contributions}
F.N.M.F., L.C.C., F.R.B, and D.M.Z conceived the project and experiments. F.N.M.F. fabricated the device. A.L. and E.P.A.M.B. synthesized the nanowire. F.N.M.F., L.C.C., O.A.H.M, F.R.B, and D.M.Z. performed the experiments. F.N.M.F., L.C.C., F.R.B., and D.M.Z analyzed the measurements and wrote the manuscript with input from all authors.

\section*{Competing interests}
The authors declare no competing interests.

\section*{Additional information}
Supplementary information is available in the online version of the paper. Reprints and permission information is available online at www.nature.com/reprints. Correspondence and requests for materials should be addressed to F.R.B. and D.M.Z.

\end{document}